# Pellet fuelling with edge-localised modes controlled by external magnetic perturbations in MAST

M Valovič, L Garzotti, C Gurl, A Kirk, D Dunai[1], A R Field, I Lupelli, G Naylor, A Thornton and the MAST team

*CCFE, Culham Science Centre, Abingdon, OX14 3DB, UK*
[1]*Wigner Research Centre for Physics, HAS, Budapest, Hungary*

E-mail: martin.valovic@ccfe.ac.uk

The fuelling of plasmas by shallow frozen pellets with simultaneous mitigation of edge-localised modes (ELM) by external magnetic perturbation is demonstrated on the MAST tokamak. Post-pellet particle loss is dominated by ELMs and inter-ELM gas fuelling. It is shown that the size of post-pellet ELMs can be controlled by external magnetic perturbations. Post-pellet ELMs remove particles from the large part of pellet deposition zone including the area with positive density gradient. The mechanism explaining this peculiarity of particle loss is suggested.

## 1. Introduction

The operation of tokamak fusion reactors depend on adequate control of plasma density and isotope mixture. Due to the high temperature and density in burning fusion plasmas conventional fuelling by gas puffing and wall recycling is predicted to be ineffective. The only candidate so far is the injection of frozen hydrogen pellets and such technique is considered for ITER fusion reactor [1]. Due to the technical limitations of pellet injectors, however, even this method of fuelling is limited to the plasma periphery, i.e. outer 20% of plasma cross section in ITER. As a consequence only a few % of the fuel injected to vessel is burned in the plasma while the rest is pumped out without reacting. Such a high fuel throughput has to be accommodated into the design of outer fuel loop systems such as pumping, tritium recovery, tritium breeding, storage and tritium environmental barriers. From the point of view of outer fuel loop one would aim for the smallest throughput (higher burn-up) as possible. Some studies suggest that the burn-up fraction should be kept above $2-5\%$ in order to guarantee self-sustained fusion reactor operation [2].

These constraints underline the importance of understanding and optimisation of the inner fuel loop which is controlled by plasma physics. The main parameter which determines the fuel throughput is the life-time of pellet-deposited material in the plasma, so called pellet retention time $\tau_{pel}$. Pellet retention time encapsulates complex physics of pellet deposition [3] and particle transport in the pellet deposition zone. The depth of pellet deposition can be improved by launching pellets from high field side of the plasma [4, 5] but otherwise it is limited by injector and launching technology such as pellet speed and pellet size. Therefore the main parameter left for optimisation is the level of particle transport in the pellet deposition zone.

Particle transport in the periphery is however subject to a number of constraints. Reduction of the particle transport by a strong H-mode transport barrier results in large ELMs which are not compatible with presently available materials for the first wall and divertor [6]. Therefore ELM size has to be actively reduced and one of the techniques envisaged in ITER is the application of resonant magnetic perturbations (RMP) [7, 8, 9].The result of these perturbation is the increase of particle transport at the plasma periphery and thus increasing the pellet particle throughput required to keep the plasma density at nominal value. Experiments with simultaneous pellet fuelling and ELM mitigation are rare and results are mixed. Some data are favourable and fuelling pellets do not affect the ELM mitigation [10, 11, 12], as it is assumed by ITER. In some experiments, however, fuelling pellets are followed by large ELMs [12, 13] which can promptly remove pellet particles what would be highly unfavourable for ITER.

The present paper reports on further expansion of the dataset of simultaneous pellet fuelling and ELM mitigation in the MAST tokamak. In particular we focus on the effects of pellets on ELM mitigation. This study thus complements the micro-stability survey in pellet deposition zone [14] which addresses the inter-ELM transport.

## 2. Experimental setup

Experiments were performed in the MAST spherical tokamak and the geometry of the experiment is shown in figure 1a. Deuterium plasma with single lower null divertor is heated by neutral beams injected tangentially in the direction of plasma current. Beams are injected horizontally in the $z=0$ plane. Deuterium pellets (diameter ~1.2mm, velocity ~ 400m/s) are injected into the plasma from top-high-field side. The choice of tangential pellet impact angle is deliberate in order to keep pellet deposition shallow so that it mimics the ITER situation. ELMs are controlled by a single row of 12 resonant magnetic perturbation coils (RMPs) with dominant toroidal mode number $n=6$. Deuterium gas is injected from multiple locations and the density of neutrals is measured by fast ion gauge at outboard mid-plane.

Plasma density is measured by a 8-laser Thomson scattering system triggered by the pellet in the flight tube with controlled delay. Line integral density is measured by interferometer with a high temporal resolution. Both Thomson scattering and interferometer measure the density along the horizontal path at vacuum vessel mid-plane $z=0$. Beam Emission Spectroscopy (BES) measures the density perturbation along the neutral beam with 2cm spatial resolution using 4x8 detector array.

## 3. Post-pellet particle loss

Figure 1b shows the typical time traces for plasma with ELM mitigation. After the application of a RMP current the density drops which is known as a density pump-out effect. In the case in figure 1b the density pump-out is partially compensated by gas fuelling leading to an increase of neutral gas density at the outboard mid-plane from $1.0 \times 10^{18} D_2 / m^3$ to $1.6 \times 10^{18} D_2 / m^3$ during the interval from 0.355s to 0.395s. As a result of these two actuators the ELM frequency increases by a factor of 2-3 and the ELM size drops by a factor of 2. After the application of RMP the neutron rate decreases by 12% which is approximately consistent with the drop of core electron temperature by 13%. Note however that due to the vertical plasma displacement the Thomson scattering data are available only for $\rho = \sqrt{\psi_N} > 0.4$, where $\psi_N$ is the normalised poloidal magnetic flux.

The plasma is refuelled by pellets which have approximately the same size as the pump-out effect. The post pellet density decay has a characteristic time constant of $\tau^*_{pel} \sim 15ms$ as seen in figure 1b on interferometer signal. It is seen that the decay time is not only the result of post pellet ELM loss but also the significant inter-ELM re-fuelling by gas. Detailed inspection of the interferometer signal shows that the inter-ELM refuelling restores roughly 60% of the ELM loss. Inter-ELM gas fuelling is somewhat stronger after the pellet compared to the pre-pellet phase because of the 30% increase of neutral gas density after the pellet. As a result the net pellet retention time $\tau_{pel}$ (i.e. without gas refuelling, as expected on ITER) is significantly shorter than the post-pellet decay time $\tau^*_{pel}$ and can be estimated as $\tau_{pel} \sim 0.4 \tau^*_{pel}$. It is also seen that without the gas refuelling in between ELMs the materials deposited by single pellet would be removed by 2-3 ELMs.

For extrapolation purposes it is useful to compare the pellet retention time with energy confinement time. Here we define the global energy confinement time as $\tau_{E,tot} = W_{mhd} / P_{div}$ where $W_{mhd}$ is the energy content from equilibrium reconstruction and $P_{div}$ is the power to divertor as measured by an infrared camera. At 0.52s we find $\tau_{E,tot} \sim 34\,\text{ms}$. Note that the standard method of validation of energy confinement time from kinetic profiles and neutron emission is impossible because the temperature and density are not measured up to the plasma core due aforementioned shift of plasma position. Taking this value for $\tau_{E,tot}$ the normalised pellet retention time can be estimated as $\tau_{pel} / \tau_{E,tot} \sim 0.17$. This value is comparable to the values reported previously on MAST for shallow deposition at $\rho = \sqrt{\psi_N} = 0.75$ [3]. It has to be noted however that the database of pellet retention time for shallow pellets is very small and clearly more data are needed for reliable prediction to ITER.

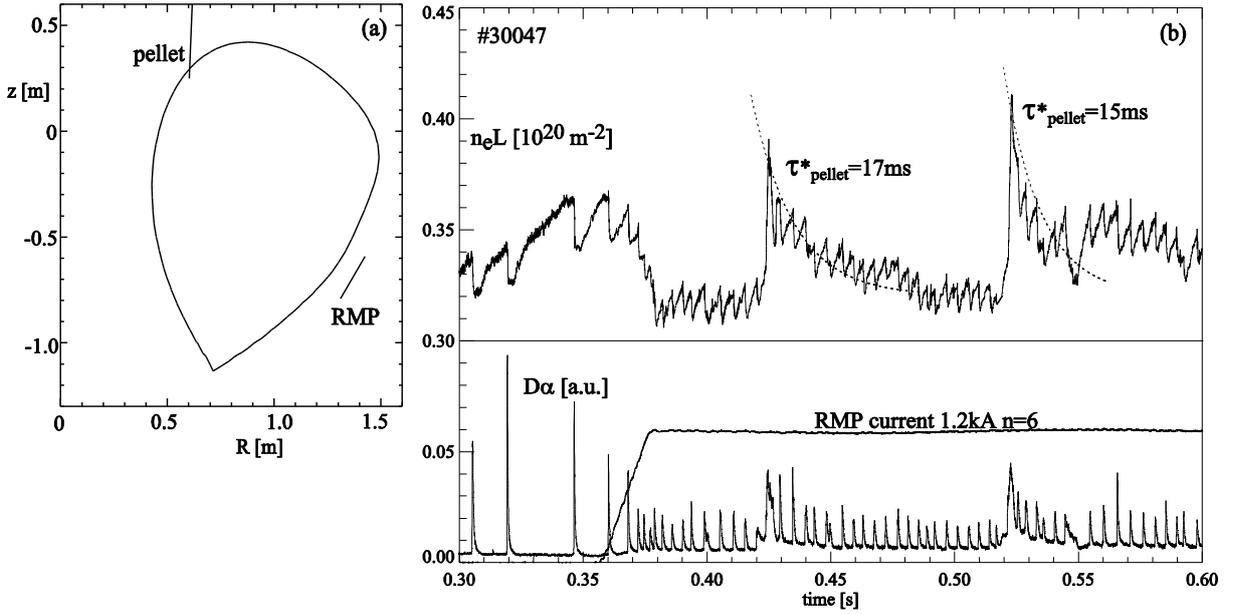

**Figure 1.** (a) the shape of the last closed magnetic surface and geometry of the experiment. (b) traces of line integrated density $n_e L$, RMP current and $D_\alpha$ emission for shot number 30047. Plasma current $I_p = 0.42MA$ and toroidal field at geometric axis $R_{geo} = 0.94m$ is $B_T = 0.43T$.

It is seen from figure 1 that the particle loss after pellet is almost entirely caused by ELMs or intermittent L-modes (compound ELMs). As mentioned above the main assumption of ITER pellet fuelling is that the pellets do not affect the ELMs or in other words that the post and pre-pellet ELMs are the same and both respond equally to the mitigation by RMPs. To test this assumption we have created a dataset in which for each pellet we have measured the relative density loss during two ELMs, one just before and one just after the pellet. The relative loss is measured by fast interferometer signal $n_e L$ and the relative loss is defined as: $\delta N / N = -1 + n_e L(\text{pre-ELM}) / n_e L(\text{post-ELM})$. The data is extracted from the raw MAST dataset automatically by a single batch process with time averaging over 0.1ms but the shots and times are selected manually. The dataset contains plasmas with variable level of ELM mitigation including reference shots without RMP. The dataset has been narrowed to plasmas only with conventional ELMs before the pellet, i.e. there are no data with compound ELMs before the pellet. Concerning the post-pellet ELMs we allow both conventional and compound ELMs in the dataset but these two groups are discussed separately. Finally we included only pellets that are smaller than 35% of the plasma content.

The dataset described above is shown in figure 2. It is seen that for conventional post pellet ELMs (full symbols in figure 2) their relative size $\delta N / N_{postpel}$ is correlated with the size of ELMs just before the pellet, $\delta N / N_{prepel}$. The Pearson correlation coefficient between these two variables is significant $R = 0.82$. This suggests that the size of post-pellet ELMs is also controlled by RMPs and not only by pellet size. This can be confirmed by the observations that the correlation between the pellet size and the size of post pellet ELMs is small ($R = 0.29$) as seen in the insert plot in figure 2. Nevertheless the pellet seems to affect the size of the post pellet ELMs as $\delta N / N_{postpel}$ is on average larger by a factor of 1.6 than the corresponding pre-pellet value $\delta N / N_{prepel}$. Note however that these factors are determined from line integrals and thus depend on actual changes of density profiles. The footprints of ELMs on density profiles will be discussed in section 4.

It is seen in figure 2 that the size of ELMs mitigated by RMPs have some lower limit at $\delta N / N_{prepel} \sim 3\%$. At this lower end of our dataset we also detect the existence of post-pellet compound ELMs, marked by open symbols in figure 2. For these ELMs the loss can be up to 3 times larger than for conventional ELMs. Such unfavourable cases were described previously [12] and it was shown that the post pellet particle loss rate could be up to 5 times higher compared to conventional ELMs. The reasons for triggering of compound ELMs by pellet are not yet well documented. The first possibility is that the RMP perturbation is too strong. In our broader database we have cases where RMP itself can trigger H-L transition even without the pellet but this is not discussed in this paper. The second reason could be that the pellet is too large. Indeed closer inspection of the scatter plot in the insert of figure 2 shows that the compound ELMs have a tendency to occur after larger pellets. These observations raise the question what is the maximum RMP current at which the post pellet compound ELMs can be avoided and how this threshold depends on plasma parameters.

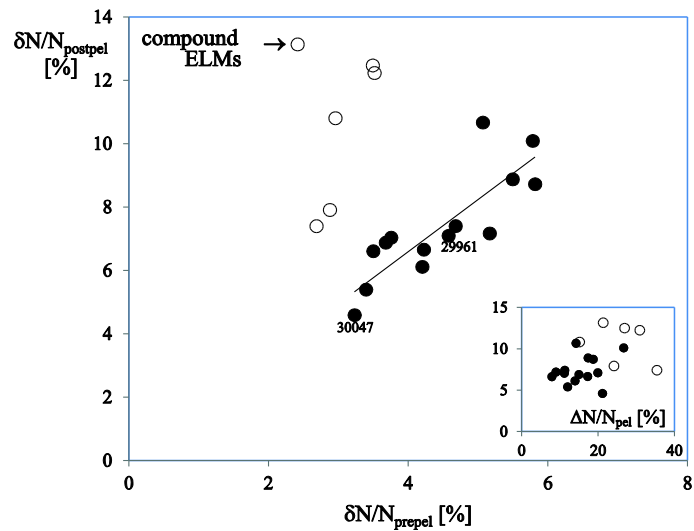

**Figure 2.** Comparison of relative particle loss for pre and post- pellet ELMs. $\delta N / N_{prepel}$ is the relative loss for ELM just before the pellet and $\delta N / N_{postpel}$ is the relative loss for ELM just after of the same pellet. Full symbols are for conventional post-pellet ELMs and open symbols are for compound post-pellet ELMs. The insert compares $\delta N / N_{postpel}$ with relative pellet size $\delta N / N_{pel}$.

## 4. Character of post-pellet ELM loss

To get insight into the mechanism of particle loss during post-pellet ELM one has to analyse the density profiles with high spatial and temporal resolution. Figure 3 shows the changes of density profiles due to pre and post-pellet ELMs. Two cases are shown: one with weak and one with strong ELM mitigation. These cases are marked by shot numbers in figure 2. The panels at the right of the figure 3 show the position of laser pulses of Thomson scattering relative to the interferometer signal. Comparison of pre and post-pellet profiles show that pellets in both cases are depositing material around the normalised minor radius of $\rho \sim 0.8$ which is similar to the situation predicted for ITER. It is seen that with increasing ELM mitigation the size of pre and post-pellet ELMs decreases, in line with global data in figure 2.

Inspection of footprints of density loss by ELMs shows that the affected area spans beyond $\rho \sim 0.8$ and its size is about the same for pre and post-pellet ELMs (figure 3). There is no significant change of density deeper into the plasma so that ELMs remove particles mainly outwards. The effect of RMPs is mostly the change of the amplitude of the density drop rather than the size of the footprint. The main difference in footprints is between pre and post pellet ELMs. While the pre-pellet ELMs act on flat density profiles, the post-pellet ELMs remove material even from the zone with inverted density gradient. This peculiarity of transporting plasma against the density gradient suggests that diffusion is not the dominant particle loss mechanism. In the next paragraph we propose a possible explanation.

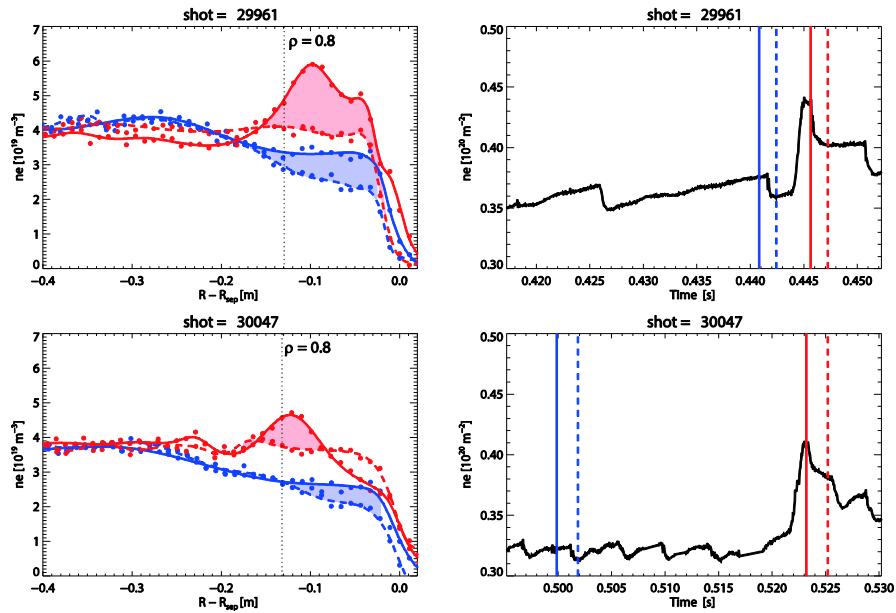

**Figure 3.** Density loss due to pre (blue) and post (red) pellet ELMs. Left column show the density profiles and right column shows the timing of profile measurement. Two plasmas are compared: top row is with weak ELM mitigation, lower row with strong mitigation. Vertical lines on profiles show location of magnetic surfaces with $\rho = 0.8$ at 0.4454s for shot 29961 and at 0.5232 for shot 30047.

To shed light on the character of post-pellet ELM loss we have inspected the data from BES diagnostics for the shot 29961 shown in figure 3 top row. This analysis is summarised in figure 4. The top panel shows the expanded temporal evolutions of the interferometer signal, $D_\alpha$ emission and one BES channel which corresponds to the normalised minor radius of $\rho \sim 0.8$ as reconstructed just before the ELM. It is seen that during the drop of line integral density the BES signal shows the perturbations of about $\pm 10\%$. This is confirmed by an analysis of all BES signals. The lower panel shows a snapshot of the whole 2D BES image taken at the time towards the end of the period of ELM density loss. The quantity plotted represents the relative perturbation of electron density $\delta n_e/\langle n_e \rangle$, where $\langle n_e \rangle$ is the temporal average of particular BES signal over 1ms. The image reveals that in the zone $\rho > 0.7$ the density is significantly perturbed up to $\delta n_e/\langle n_e \rangle \sim -0.15$. The surfaces of constant density perturbations are not aligned with pre-ELM magnetic surfaces and characteristic size of the perturbation is about $2L \sim 9 \text{cm}$. This length is taken along the line connecting the surfaces with $\delta n_e/\langle n_e \rangle \sim -0.15$ (dotted line in figure 4). Let us assume that these density perturbations are also accompanied by perturbations of electrostatic potential $\delta \varphi$. If we now estimate the amplitude of normalised electrostatic potential $e\delta\varphi/T_e$ by the size of relative density perturbation $\delta n_e/\langle n_e \rangle$ we get the $\mathbf{E} \times \mathbf{B}$ drift velocity of the order of $\delta v_\perp \sim \delta E/B_T \sim 2 km/s$. Here, $T_e(\rho = 0.85) \sim 200 eV$, $\delta E \sim \delta\varphi/L$ and $B_T(1.3m) = 0.31T$. Such a velocity would displace the plasma along the structure's size $\sim 2L$ in about $\tau_\perp \sim 2L/\delta v_\perp \sim 43 \mu s$. This timescale is comparable with the ELM duration and thus such plasma flow pattern is capable of transporting the particles from the pellet deposition zone to the plasma edge even against the density gradient. Note that this case the size of the structure is of the same order of magnitude as the density gradient scale-length. Therefore the particle flow is not described by a conventional picture of turbulent transport in which the spatial size of density perturbations are much smaller than the scale-length of main average density profile.

Finally we are aware that the above evaluation of the electric field from the density perturbations cannot be taken too strictly because the perfect alignment of surfaces of electric filed with surfaces of constant density perturbations will not result in overall particle transport. This evaluation should be considered as an order of magnitude estimate and more detailed model and simulation are required to quantify the plasma flow. For a discussion of departure from the Boltzmann relation between density and electrostatic potential see [15].

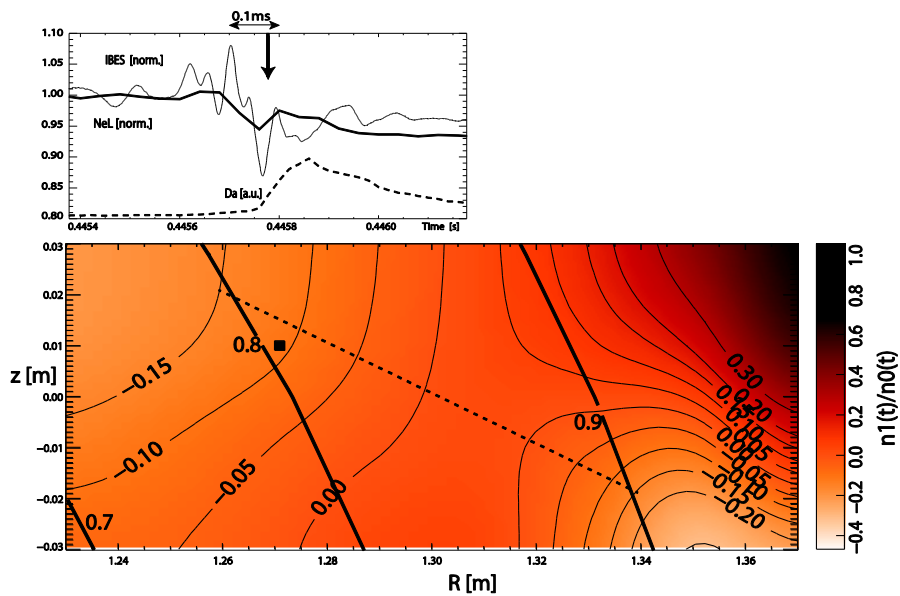

**Figure 4.** Detail of the particle loss during post-pellet ELM for shot 29961. Top panel: temporal traces of line integrated density $n_e L$, BES signal from detector localised at $\rho \sim 0.8$ and $D_\alpha$ emission. BES and line integral density are normalised to pre-ELM values. Lower panel: BES 2D image taken at the time shown by arrow on the top panel. Solid lines are surfaces of constant $\delta n_e/\langle n_e \rangle$. Thick solid lines are the poloidal flux surfaces of $\rho = const$ evaluated just before the ELM at 0.4454s. The square symbol shows the position of detector of BES signal on the top panel. The dotted line shows path along which the characteristic size of the structure $2L$ was evaluated.

## 5. Conclusions

The paper reports on new data from the MAST tokamak in which plasma was fuelled by shallow high field side pellets simultaneously with ELM mitigation with RMP fields - the two key actuators for density control in ITER. In our experiment post-pellet particle loss is dominated by ELMs. Inter-ELM losses are small and masked by significant gas fuelling.

Our data show that pellet fuelling and RMP ELM mitigation can be compatible in the sense that: (1) the size of post-pellet ELMs responds to ELM mitigation and (2) the post-pellet ELMs are not significantly larger than pre-pellet ELMs. These favourable observations are however muted by the fact that the relative size of post-pellet ELMs is still quite large where 2 - 3 ELMs are sufficient to remove the material deposited by a single pellet - a much smaller number of ELMs than expected in ITER.

A detailed inspection of post-pellet ELMs shows that the ELM related density drop can cover the whole pellet deposition zone and is consistent with outward particle flow. The ELM affected area also includes the zone with inverted density gradient by the pellet raising the question about the mechanism of outward particle flow against the positive density gradient. Data from BES diagnostics during post-pellet ELMs reveal the existence of a sizeable structure suggesting that the large scale convection patterns can explain this peculiarity. This analysis was performed only for post pellet ELMs with RMPs and we have not attempted to generalise these finding further.

The experiments on shallow pellet fuelling simultaneously with ELM mitigation are still rare. Clearly more data are needed to demonstrate that these two actuators will be fully capable to control the plasma density and isotope content in ITER under all possible conditions.

## Acknowledgements

This project has received funding from the European Union's Horizon 2020 research and innovation programme under grant agreement number 633053 and from the RCUK Energy Programme [grant number EP/I501045]. To obtain further information on the data and models underlying this paper please contact PublicationsManager@ccfe.ac.uk. Authors would like to thanks to Dr A. Thyagaraja for valuable discussion and Dr Brian Lloyd careful reading of the manuscript.

## References

[1] Kukushkin A S, Polevoi A R *et al* 2011 *Journal of Nuclear Materials* **415** S497
[2] Sawan M E and Abdou Z A 2006 *Fusion Engineering and Design* **81** 1131
[3] Valovič M *et al* 2008 *Nucl. Fusion* **48** 075006
[4] Lang P T *et al* 1997 *Phys. Rev. Lett.* **79** 1487
[5] Baylor L *et al* 2000 *Phys. Plasmas* **8** 1878
[6] Loarte A *et al* 2014 Nucl. Fusion **54** 033007
[7] Evans T *et al* 2008 *Nucl. Fusion* **48** 0244002
[8] Kirk A *et al* 2012 *Phys. Rev. Lett.* **108** 25503
[9] Suttrop W *et al* 2011 *Phys. Rev. Lett.* **106** 225004
[10] Lang P T *et al* 2012 *Nucl. Fusion* **52** 024002
[11] Liang Y *et al* 2010 *Plasma Fusion Res*. **5** S2018
[12] Valovič M *et al* 2013 *Plasma Phys. Control. Fusion* **55** 025009
[13] Baylor L et al 2008 Proc. 35th EPS Conf. on Plasma Physics (Hersonissos, Greece 2008) vol 32D (ECA) P 4.098
[14] Garzotti L *et al* 2014 *Plasma Phys. Control. Fusion* **56** 035004
[15] Thyagaraja A and Haas F A 1993 *Physica Scripta* **47** 266